
\input epsf

\ifx\epsffile\undefined\message{(FIGURES WILL BE IGNORED)}
\def\insertfig#1#2{}
\else\message{(FIGURES WILL BE INCLUDED)}
\def\insertfig#1#2{{{\baselineskip=4pt
\midinsert\centerline{\epsfxsize=\hsize\epsffile{#2}}{{
\centerline{#1}}}\medskip\endinsert}}}
\fi

\input harvmac

%
%
%
%
%
\ifx\answ\bigans
\else
\output={

\almostshipout{\leftline{\vbox{\pagebody\makefootline}}}\advancepageno

}
\fi
%
%
%

%
%

%
%
\def\UCSD#1#2{\noindent#1\hfill #2%
\bigskip\supereject\global\hsize=\hsbody%
\footline={\hss\tenrm\folio\hss}}
%
%
\def\abstract#1{\centerline{\bf Abstract}\nobreak\medskip\nobreak\par
#1}
%
%
%
%
\edef\tfontsize{ scaled\magstep3}
 \tfontsize  \tfontsize
\font\titlermss=cmr5 \tfontsize \font\titlei=cmmi10 \tfontsize
\font\titleis=cmmi7 \tfontsize \font\titleiss=cmmi5 \tfontsize
\font\titlesy=cmsy10 \tfontsize \font\titlesys=cmsy7 \tfontsize
\font\titlesyss=cmsy5 \tfontsize  \tfontsize
\skewchar\titlei='177 \skewchar\titleis='177 \skewchar\titleiss='177
\skewchar\titlesy='60 \skewchar\titlesys='60 \skewchar\titlesyss='60
\scriptscriptfont0=\titlermss
\scriptscriptfont1=\titleiss
\scriptscriptfont2=\titlesyss
%
%
%

%
\def\inv{^{\raise.15ex\hbox{${\scriptscriptstyle -}$}\kern-.05em 1}}
\def\lbar{{\lower.35ex\hbox{$\mathchar'26$}\mkern-10mu\lambda}}

%
%
%
%
\def\slash#1{\rlap{$#1$}/} 
\def\dsl{\,\raise.15ex\hbox{/}\mkern-13.5mu D} 

\def\delsl{\raise.15ex\hbox{/}\kern-.57em\partial}
\def\Ksl{\hbox{/\kern-.6000em\rm K}}
\def\Asl{\hbox{/\kern-.6500em \rm A}}
\def\Dsl{\hbox{/\kern-.6000em\rm D}} 
\def\Qsl{\hbox{/\kern-.6000em\rm Q}}
\def\gradsl{\hbox{/\kern-.6500em$\nabla$}}
%
%
\def\lspace{\ifx\answ\bigans{}\else\qquad\fi}
\def\lbspace{\ifx\answ\bigans{}\else\hskip-.2in\fi} 
%
%
\def\boxeqn#1{\vcenter{\vbox{\hrule\hbox{\vrule\kern3pt\vbox{\kern3pt
        \hbox{${\displaystyle #1}$}\kern3pt}\kern3pt\vrule}\hrule}}}
%
%
\def\mbox#1#2{\vcenter{\hrule \hbox{\vrule height#2in
\kern#1in \vrule} \hrule}}
%
%
%
%

  \def\CO{{\cal O}}

%
%
%
%
%

%

\def\bar#1{\overline{#1}}

\def\darr#1{\raise1.5ex\hbox{$\leftrightarrow$}\mkern-16.5mu #1}

%
%
\def\frac#1#2{{\textstyle{#1\over #2}}} 
%
%
%
%

\def\GeV{{\rm GeV}}

%
%
%
%

%
%
\def\ltap{\ \raise.3ex\hbox{$<$\kern-.75em\lower1ex\hbox{$\sim$}}\ }
\def\gtap{\ \raise.3ex\hbox{$>$\kern-.75em\lower1ex\hbox{$\sim$}}\ }
\def\gl{\ \raise.5ex\hbox{$>$}\kern-.8em\lower.5ex\hbox{$<$}\ }
\def\roughly#1{\raise.3ex\hbox{$#1$\kern-.75em\lower1ex\hbox{$\sim$}}}

%
%

%

%
\def\np#1#2#3{{Nucl. Phys. } B{#1} (#2) #3}
\def\pl#1#2#3{{Phys. Lett. } {#1}B (#2) #3}
\def\prl#1#2#3{{Phys. Rev. Lett. } {#1} (#2) #3}

\relax

\def\hbar{\bar h_Q}

\def\qsl{\hbox{/\kern-.5600em {$q$}}}
\def\ksl{\hbox{/\kern-.5600em {$k$}}}

\def\({\left(}
\def\){\right)}

\def\OMIT#1{}
\def\frac#1#2{{#1\over#2}}

\def\etwohat{{y_d}}
\def\eonehat{{y_{\bar u}}}
\def\echat{{y_q}}
\def\mqhat{{\hat m_q}}
\def\pvec{\vec p}
\def\pchat{{\hat{p}_q}}
\def\xqone{\cos\theta_{q\bar u}}
\def\xqtwo{\cos\theta_{qd}}
\def\xonetwo{\cos\theta_{{\bar u}d}}
\def\xcos{\cos\theta}
\def\xcosqu{\cos^2\theta}
\def\xcoscu{\cos^3\theta}
\def\xcosfour{\cos^4\theta}
\def\xcosfive{\cos^5\theta}
\def\as{\alpha_s(m_b)}

\hbadness=10000

\noblackbox
\vskip 1.in
\centerline{{\titlefont{Hadronic Event Shapes in $B$ Decay}}}
\medskip
\vskip .5in
\centerline{Michael Luke${}^{a}$, Martin J.~Savage${}^{b}$
and Mark B.~Wise$^{c}$}
\medskip
{\it{
\centerline{a) Department of Physics, University of Toronto,
Toronto,
Canada M5S 1A7}
\centerline{b) Department of Physics, Carnegie Mellon
University,
Pittsburgh PA 15213}
\centerline{c) Department of Physics, California Institute
of Technology,
Pasadena, CA 91125}}}

\vskip .2in

\abstract{
We compute the differential momentum correlation function
for hadrons produced in the decay of $B$-mesons.  This measure of
hadronic event shapes tests the free $b$-quark decay picture for
nonleptonic and semileptonic decays of $B$-mesons.  Our results can also
be applied to $B_s$ and $\Lambda_b$ decay.}

\vfill
\UCSD{\vbox{
\hbox{UTPT 93-28}
\hbox{CMU-HEP 93-24}
\hbox{CALT-68-1905}
\hbox{DOE-ER/40682-49}}
}{November 1993}
\eject

Free heavy quark decay is expected to  describe accurately the
inclusive weak decays of hadrons containing a heavy quark.  There are
two physical reasons for this.
First, while the individual exclusive modes depend on the details of
hadronization,
inclusive quantities which sum over all possible decay modes are
expected to be insensitive to these details, since for heavy quarks there
are many states available to the decay products.  Second, for large
heavy quark mass $m_Q$, the heavy quark in the hadron is almost on-shell
and moving with the same four-velocity as the hadron.  It is therefore
reasonable to treat its decay as that of a free quark.

For semileptonic (and radiative) decays, this duality between
hadrons and free quarks and gluons has been shown to follow from QCD as the
leading term in an expansion in $1/m_Q$
\ref\cgg{J.~Chay, H.~Georgi and B.~Grinstein, \pl{247}{1990}{399}.}.
At leading order the differential decay rate for inclusive
$B \rightarrow X e\bar \nu_e$ decay,
$d\Gamma/dq^2 dE_e~ (q  = p_e + p_\nu)$, suitably averaged over $E_e$,
is equal to that for free $b$-quark decay.
Non-perturbative corrections, parameterising the effects of the
strong interactions
of the heavy $b$-quark with the light degrees of freedom in the
$B$-meson, first arise at order $1/m_b^2$. These have recently been
calculated
\ref\bigietala{I.I.~Bigi, N.~G.~Uraltsev and A.~I.~Vainshtein,
\pl{293}{1992}{430}.}--%
\nref\bigietalb{I.~I.~Bigi, B.~Blok, M.~Shifman, N.~G.~Uraltsev and
A.~Vainshtein,
TPI-MINN-92/67-T (1992); \prl{71}{1993}{496}.}%
\nref\bloketal{B.~Blok, L. Koyrakh, M.~Shifman and A.~Vainshtein,
NSF-ITP-93-68 (1993), hep-ph/9307247.}%
\nref\manwise{A.~V.~Manohar and M.~B.~Wise, UCSD-PTH 93-14 (1993),
hep-ph/9308246.}%
\nref\mannel{T.~Mannel, IKDA 93/16 (1993), hep-ph/9308262.}%
\ref\fls{A.~F.~Falk, M.~Luke and M.~J.~Savage, UCSD-PTH 93-23
(1993), hep-ph/9308288.}.
Duality in this instance arises because the current-current correlator
for the weak hadronic current is analytic everywhere in the complex
energy plane, except at points on the real axis corresponding to physical
intermediate states.  This allows the phase space integral over
the energy of the final hadronic state to be deformed
to lie far from the the singularities in the
decay amplitude.  Along the deformed contour perturbative QCD provides
a valid description of the decay products free of infrared
singularities.  The free quark decay picture emerges at leading order in
$1/m_b$.

For nonleptonic decays the above argument does not apply, since there is no
longer a kinematic variable that can be used to define a path of
integration to be deformed.  Nevertheless, the free quark decay picture is
physically reasonable and widely used (e.g., to extract $|V_{cb}|$ from
the $B$-meson lifetime).
Even if the free quark decay picture holds when $m_b$ is very large for
the physical value of the $b$-quark mass there is only about $m_b - m_c
\simeq 3\,\GeV$ of energy (beyond the charm quark mass) available to the
final state particles in weak $B$ decay.  Furthermore, this energy is
shared  amongst three quarks in $b \rightarrow c\bar u d$ nonleptonic
decay so that the energy that each quark carries is not very large
compared with the QCD scale.  It is important to test the free quark decay
picture in
as many ways as possible.  One test is provided by the predicted
equality of the $B^0, B^-, B_s$ and $\Lambda_b$ lifetimes and another by
the predicted semileptonic branching ratios.
In this letter, we point out that event shapes, characterized by
the momentum angular correlation distribution, may also be used to test its
validity.  We calculate this correlation function for both nonleptonic and
semileptonic $b$ decays assuming that the free quark decay picture is
valid.

Consider the weak decay of a hadron containing a $b$-quark.  We define the
following quantity, in analogy with the differential energy-energy correlation
function used to describe hadronic events at colliders \ref\BP{V.~Barger
and R.~Phillips, {\it Collider Physics}, Addison-Wesley (1987) p. 291.}:
\eqn\enen{\eqalign{ {d\Sigma\over d\cos\theta} =
{1\over \Gamma}&\left(
\sum_{i, j} \int\,dE_i \,dE_j  \ { | \pvec_i | \over m_b} \ {|
\pvec_j | \over m_b} \
{d^3\Gamma\over dE_i \ dE_j \ d\cos\theta_{ij} }\right.\cr
+&\left.\sum_i \int\,dE_i {|\pvec_i|^2\over m_b^2}
{d\Gamma\over dE_i} \delta (1-\xcos)\right).}}
$\Gamma$ is the total nonleptonic or semileptonic decay width, $m_b$
is the mass of the initial heavy $b$-quark, and
the double sum runs over all pairs of strongly interacting particles
$(i,j),~i\not= j$, in the final state separated by an angle $\theta_{ij}$,
treating $(i,j)$ and $(j,i)$ as distinct pairs.
As we have defined it, the correlation function \enen\ is insensitive
to the emission of soft and collinear gluons by the partons in the
final state, which is required for it to be free of infrared
divergences.  Consequently $d\Sigma/d\cos\theta$ at the hadron level should
be equal (up to corrections of order $1/m_b$) to its value calculated
from free $b$-quark decay.
For $b \rightarrow c$ transitions the final state contains a charmed
hadron which decays weakly.  It is important in the evaluation of (1)
that this charmed hadron be treated as a single particle.  One should
not evaluate the sums in (1) using the charmed hadron's decay products. This
complication makes $d\Sigma/d\cos\theta$ more difficult to measure.

For nonleptonic $b$ decay, at leading order in $\alpha_s$, the tree
level $b \rightarrow q \bar u d$ matrix element of the weak Hamiltonian gives
\eqn\diff{\eqalign{{d^3\Gamma\over d\echat\,d\etwohat
\,d\xqtwo} = &
{3G_F^2m_b^5\over 2\pi^3}
\left({|\pchat|\etwohat^2 (\echat - |\hat p_q|\cos \theta_{qd})
(1-\echat-\etwohat)
\over
1-\echat+|\pchat| \xqtwo }\right) \cr
&\qquad\times\delta \left(\etwohat - {1-2\echat+\mqhat^2\over 2(1-\echat+
|\pchat| \xqtwo)}\right)\cr
{d^3\Gamma\over d\echat \,d\eonehat \,d\xqone}=&
{3G_F^2m_b^5\over 2\pi^3}
\left({ |\pchat|\eonehat^2 (\echat-\mqhat^2-\eonehat\echat+\eonehat
|\pchat|\xqone)\over
1-\echat+|\pchat| \xqone}\right) \cr
&\qquad\times\delta \left(\eonehat - {1-2\echat+\mqhat^2\over
2(1-\echat+|\pchat|
\xqone)}\right)\cr
{d^3\Gamma\over d\eonehat \,d\etwohat
\,d\xonetwo } = &
{3G_F^2m_b^5\over 2\pi^3}\
\left({ \eonehat^2\etwohat^2 (1-\eonehat+\eonehat \xonetwo)   \over
1-\eonehat+\eonehat \xonetwo }\right) \cr
&\qquad\times\delta \left(\etwohat - {1-2\eonehat-\mqhat^2\over
2(1-\eonehat+
\eonehat \xonetwo)}\right)}}
\medskip\noindent
where $\echat=E_q/m_b$, $\eonehat=E_u/m_b$, $\etwohat=E_d/m_b$,
$\pchat=p_q/m_b$ and $\mqhat=m_q/m_b$.
For the case, $q=u$, the energy integral
in \enen\ may be performed analytically, and the correlation function is
found to be
\eqn\massless{\eqalign{ \left.{d\Sigma\over d\xcos}\right\vert_{m_q=0}
 = & {8\over (1-\xcos)^7}
\left[\vphantom{(1+\xcos\over 2)} (1-\xcos)(811+1163\xcos+401
\xcosqu+25\xcoscu) \ \right. \cr
& \left. +\,6 (1+\xcos)(195+171\xcos+33\xcosqu
+\xcoscu)\log{1+\xcos\over 2}\right]\cr
& +  {11\over 30}\,\delta (1-\xcos).}}
When $q = c$ the mass of the charm quark cannot be neglected and, the
phase space integration  must be done numerically.  In
\fig\nonlep{The momentum correlation function $d\Sigma/d\cos\theta$
for nonleptonic decays computed in the parton model to lowest order
in $\alpha_s$.
The solid curve corresponds to the case when all three quarks in the
final state
are massless.
The dashed curve corresponds to the case for a $b$ quark decaying to
a
charmed final state.}
the momentum correlation functions for $q=u$ (solid curve) and $q=c$
(dashed curve) are shown for $m_b = 4.8\,\GeV$ and $m_c = 1.6\,\GeV$.
In
both cases, the correlation function is strongly peaked at $\theta =
\pi$, along with the delta function at $\theta=0$.
In addition to corrections suppressed by powers of $1/m_b$, there
are computable $\CO\(\alpha_s(m_b)\)$ corrections to the results
in \nonlep.

For comparison with experiment, it is also useful to define the Fox-Wolfram
moments $H(L)$ \ref\fw{G.~Fox and S.~Wolfram, \prl{23}{1979}{1581}}
of the distribution \enen:
\eqn\fwm{H(L) = \int\,d\cos\theta\,{d\Sigma\over d\cos\theta}
P_L(\cos\theta )}
where $P_L(x)$ are Legendre polynomials.
The first few Fox-Wolfram moments $H(L)-H(0)$
(we have subtracted $H(0)$ to remove the effect of the delta function
at $\theta=0$) are presented in Table 1.  $H(1)$ is equal to zero
because of momentum conservation.  Conservation of energy implies, in
the massless $q = u$ case, that $H(0) = 1$.

We now consider inclusive semileptonic $B$ decays.
At leading order in $\alpha_s$ there is only one quark
in the final state, so the correlation function is proportional
to a delta function at $\theta=0$:
\eqn\semione{\eqalign{{d\Sigma^{(0)} \over d\xcos} =& {2\over 15}
{1-9\mqhat^2+45\mqhat^4-
45\mqhat^8+9\mqhat^{10} -\mqhat^{12}+120\mqhat^6\log\mqhat  \over
1-8\mqhat^2+8\mqhat^6-\mqhat^8-24\mqhat^4\log\mqhat}\cr
&\qquad\qquad\times\delta (1-\xcos).}}
The leading perturbative contribution to the semileptonic correlation
function away from $\cos\theta=1$ arises
from gluon bremsstrahlung at $\CO(\alpha_s)$, and may be
calculated using the results of \ref\gpr{B.Guberina, R.D. Peccei and R. Ruckl,
\np{171}{1980}{333}.}.
The rate for the process $b\rightarrow qge \bar\nu_e$ is
\eqn\semirate{\Gamma = {1\over 2 m_b} {1\over (2\pi)^8}
\int {d^3 p_c\over 2E_c}
{d^3 k\over 2E_g}
{d^3 p_e\over 2E_e} {d^3 p_\nu\over 2E_\nu}
\delta^4( p_b-p_c-p_e-p_\nu-k)
{1\over 2}|{\cal M}|^2,}
where the spin averaged matrix element for is
\eqn\semimat{\eqalign{  &
{1\over 2}|{\cal M}|^2 = G_F^2g_s^2
Tr \left[ \slash{p}_e \gamma_\mu \slash{p}_\nu
\gamma_\beta (1-\gamma_5) \right]\cdot \cr
& \ Tr \left[ \slash{p}_c
\left( {2p_c^\lambda+\gamma^\lambda\slash{k} \over p_c\cdot k}
\gamma_\mu -
\gamma_\mu {2p_b^\lambda-\slash{k}\gamma^\lambda \over
p_b\cdot k} \right)
\slash{p_b}\
\left( \gamma_\beta {2p_{c\lambda} - \slash{k}
\gamma_\lambda \over p_c\cdot k} -
{2p_{b\lambda}-\gamma_\lambda\slash{k}\over p_b\cdot k}
\gamma_\beta \right)
\right]\,\, , } }
and $k$ is the gluon four-momentum.
The integration of the lepton phase space gives
\eqn\lep{\int {d^3p_e\over 2E_e}  {d^3 p_\nu\over 2E_\nu}
Tr \left[ \slash{p}_e \gamma_\mu \slash{p}_\nu \gamma_\beta
(1-\gamma_5) \right]
= {2\pi\over 3}\left[ P_\mu P_\beta - g_{\mu\beta}
P^2\right]\ \ \ \ ,}
where $P=p_e+p_\nu$.
Once again, the integration can be performed analytically when
the final quark is massless, and the $\CO(\alpha_s)$ contribution
to the correlation function is found to be
\eqn\semimless{ \eqalign{ &\left.{d\Sigma^{(1)} \over d\xcos}
\right\vert_{m_{q}=0} =
-{4\alpha_s\over 3\pi} \left({(83-5\xcos)(1+\xcos)^3\over
3(1-\xcos)^7} \log{1+\xcos\over 2}  \right. \cr
&\qquad +\left. {6875+17373\xcos+12168\xcosqu+1468\xcoscu-507\xcosfour+
63\xcosfive\over 360(1-\xcos)^6} \right)\,\, ,}}
where we have neglected the $\CO(\alpha_s)$ contribution to the delta
function.  The contribution of $d\Sigma^{(1)}/d\cos\theta$, in eq. (9),
to $H(0)$ diverges.  This divergence is cancelled by contribution of
the order $\alpha_s$ part of the delta function (that we neglected) to
$H(0)$.  For a massive $c$ quark in the final
state, we have performed the phase space integration numerically.
The results are plotted for $q=u$ and $q=c$ in
\fig\semiless{The momentum correlation function for
(a) $b\rightarrow uge \bar\nu_e$ and (b) $b\rightarrow cge \bar\nu_e$
as a function of
$\cos\theta$, omitting the normalization factor $4\alpha_s/3\pi$.
The dominant contribution to the delta function at $\theta=0$ comes from
the tree-level graph; away from $\theta=0$ the distribution arises from
the parton-level decay $b\rightarrow qge \bar\nu_e$.}.
Unfortunately, for both $q=u$ and $q=c$ there is only a small
contribution to the correlation function away from the forward
direction, making these distributions extremely difficult to measure.

The first few Fox-Wolfram moments $H(L)-H(0)$ for semileptonic $b$ decays
are presented in Table 2.  Note that these moments are much larger for
$b \rightarrow u$ transitions than $b \rightarrow c$ transitions.  The
large charm quark mass suppresses the amplitude for gluon bremsstrahlung.

Clearly the results given in this paper also hold for $\Lambda_b$ and
$B_s$ decay.  For the $q = u$ case, where the final quark is massless,
our
results can be applied to charm decay.   However, the difference in lifetimes
between the $D^+$ and $D^0$ is a clear sign that the free quark decay
picture breaks down badly for nonleptonic charm decays, and so we do not
expect our results to be useful in that case.

Finally, we note that if the predictions for $b$ decay event shapes
that we have made
disagree with experiment, it will not provide conclusive evidence
that the free quark decay picture fails for the total nonleptonic
decay width.  It is possible that corrections suppressed by powers of
$1/m_b$ are larger for event shapes than for the nonleptonic width.

\bigskip

We thank H.D. Politzer and L. Wolfenstein for useful discussions.
This research was supported in part by
the Department of Energy under contract DE--FG02--91ER40682 and
DE-FG03-92-ER40701.

\listrefs
\listfigs
\vfill\eject
\input tables
\begintable
{}~~|  \multispan{2} ~~$H(L)-H(0)$ ~~ \cr
{}~~$L$~~|~~~~~$m_q=0$~~~~|~~~$m_q=1.6\,\GeV$~~~\cr
$1$ | $-1.0$ | $-0.72$\crnorule
$2$ | $-0.44$ | $-0.31$\crnorule
$3$ | $-0.74$ | $-0.54$\crnorule
$4$ | $-0.57$ | $-0.41$\crnorule
$5$ | $-0.67$ | $-0.49$\crnorule
$6$ | $-0.61$ | $-0.44$
\endtable
\medskip
\centerline{{\bf Table 1: Fox-Wolfram Moments for Nonleptonic $B$ Decay}}
\bigskip
\bigskip
\begintable
{}~~|  \multispan{2} ~~$3\pi/4\as\times\(H(L)-H(0)\)$ ~~ \cr
{}~~$L$~~|~~~~~$m_q=0$~~~~|~~~$m_q=1.6\,\GeV$~~~\cr
$1$ | $-0.15$ | $-0.028$\crnorule
$2$ | $-0.26$ | $-0.037$\crnorule
$3$ | $-0.32$ | $-0.035$\crnorule
$4$ | $-0.36$ | $-0.035$\crnorule
$5$ | $-0.40$ | $-0.034$\crnorule
$6$ | $-0.43$ | $-0.034$
\endtable
\medskip
\centerline{{\bf Table 2: Fox-Wolfram Moments for Semileptonic $B$ Decay}}
\vfill\eject
\insertfig{Figure 1 }{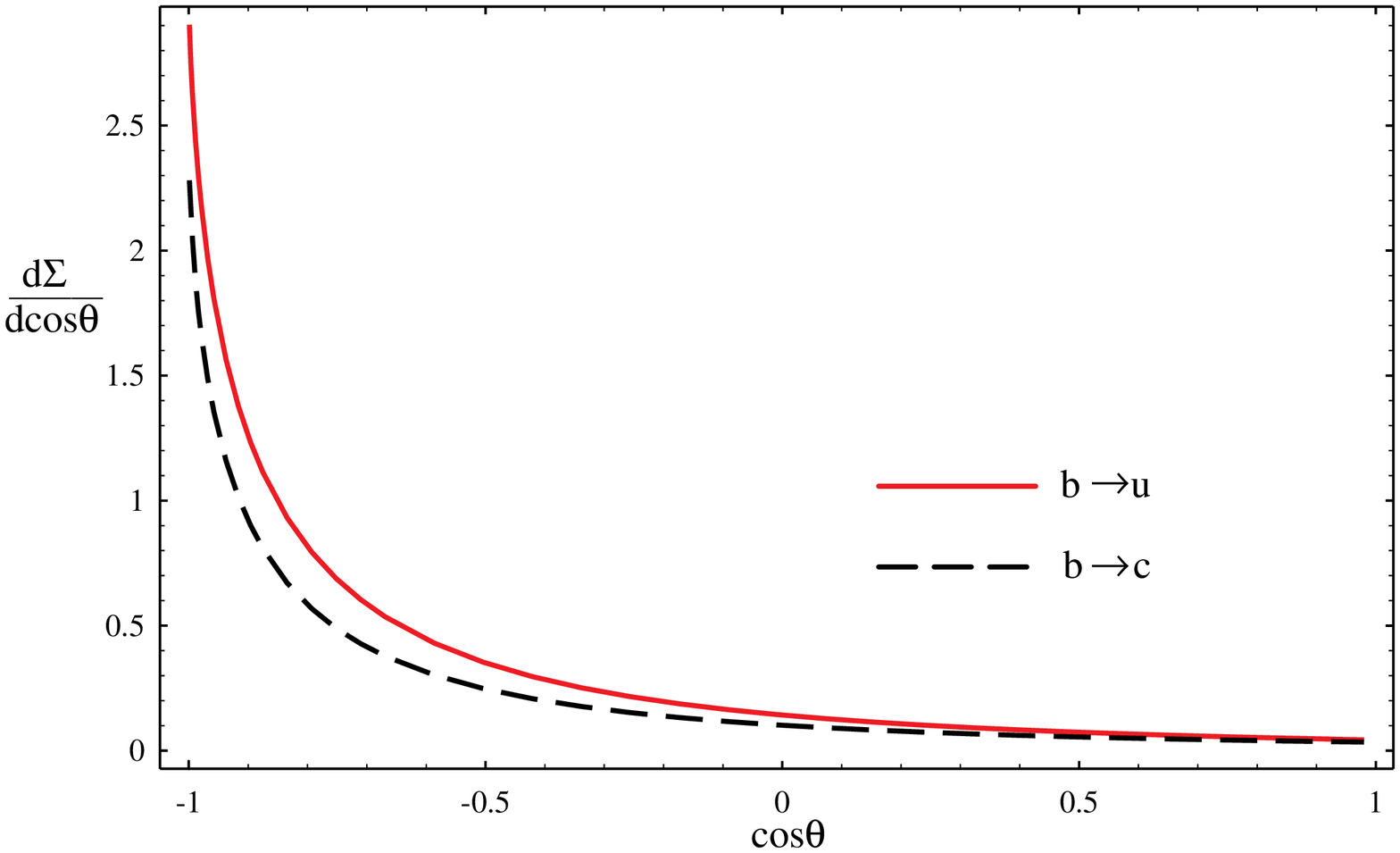}
\insertfig{Figure 2 (a) }{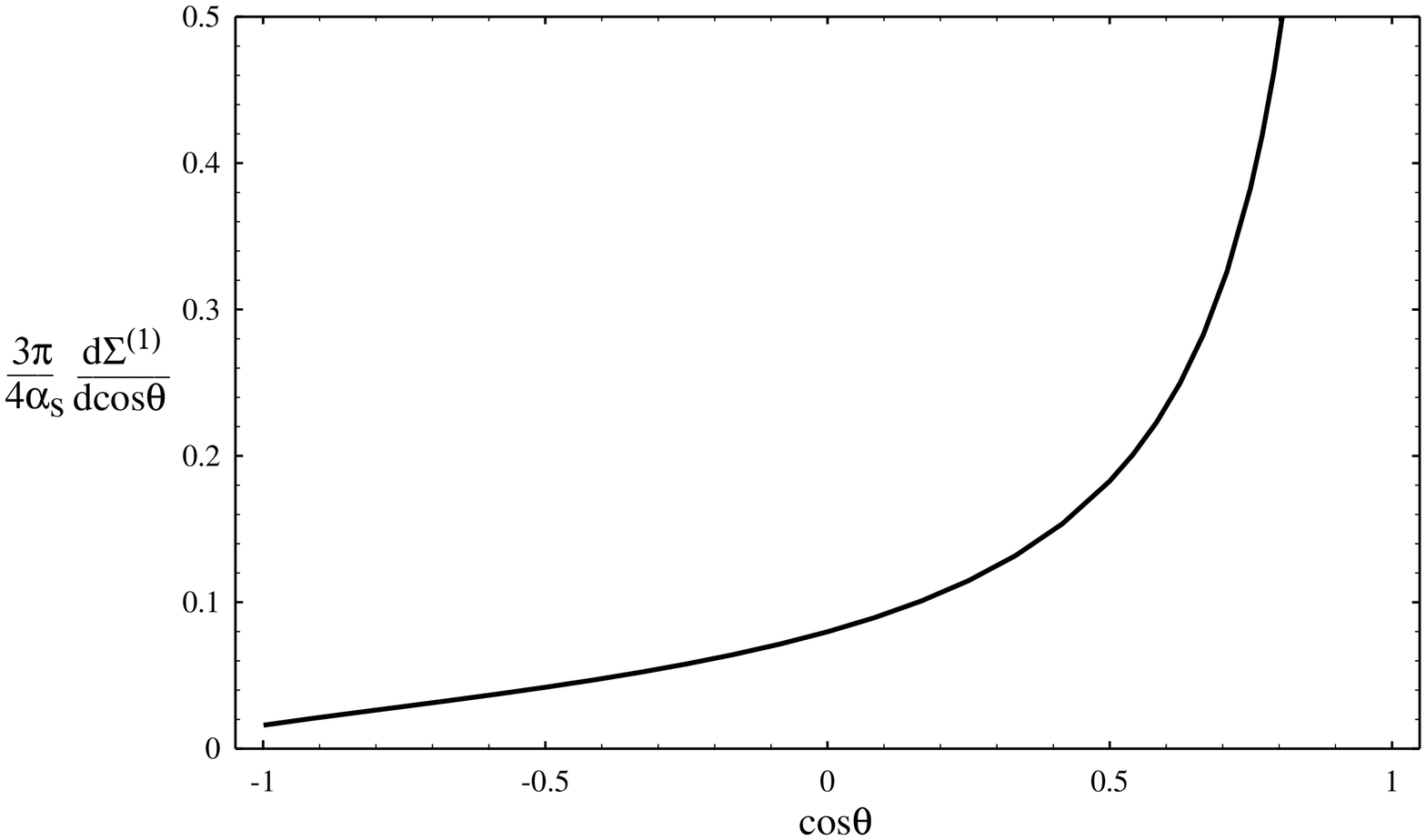}
\insertfig{Figure 2 (b) }{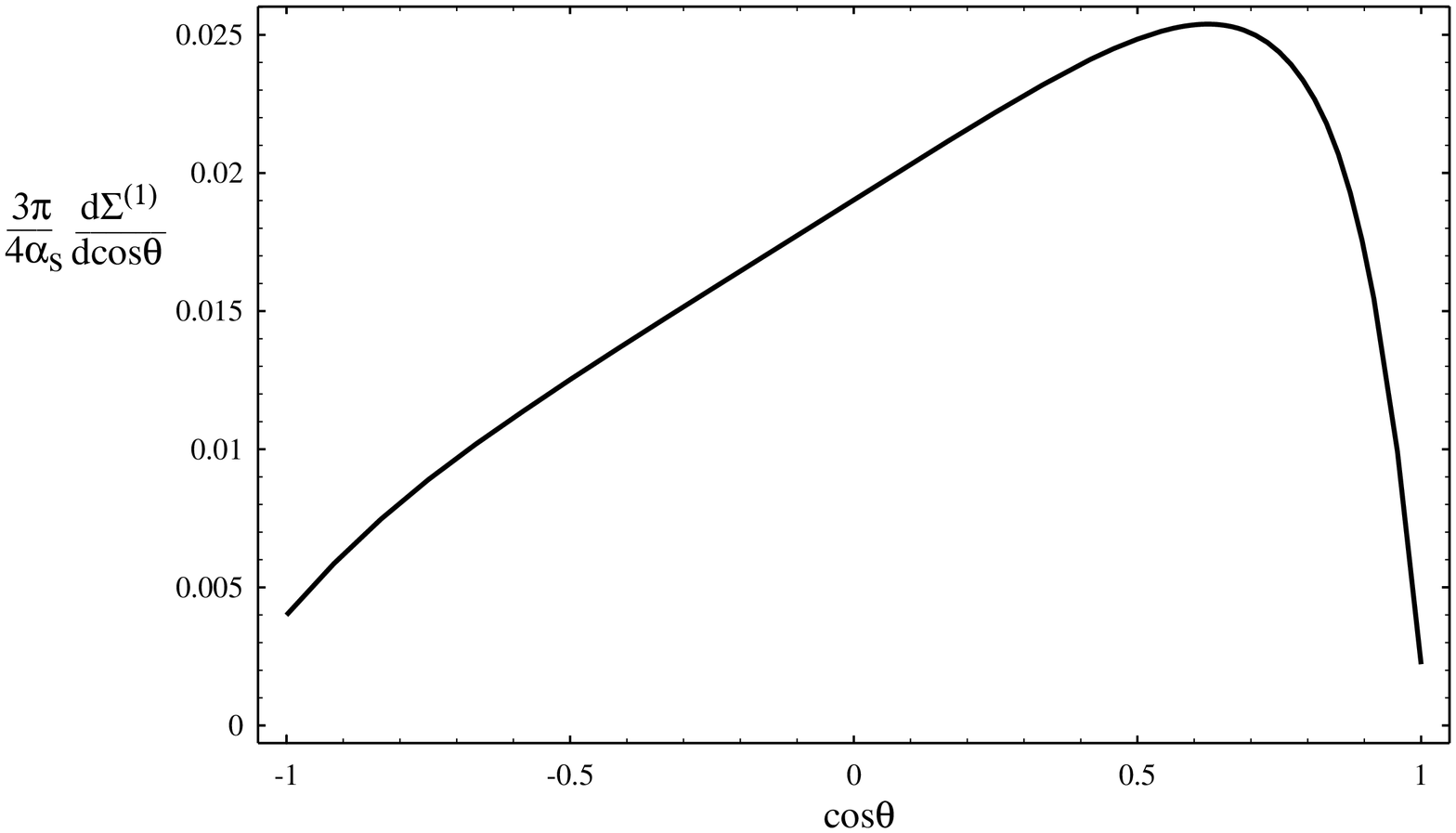}
\vfill\eject
\bye